\documentclass[a4paper,11pt]{article}

\pdfoutput=1

\usepackage{jinstpub}
\usepackage{amsmath}
\usepackage{mathtools} 
\usepackage{siunitx} 
\usepackage{placeins} 
\usepackage{upgreek} 
\usepackage[shortcuts]{extdash} 
\usepackage{booktabs} 

\graphicspath{{figures/}}
\newcommand{\fig}[1]{figure~\ref{fig:#1}} %
\newcommand{\Fig}[1]{Figure~\ref{fig:#1}} 

\newcommand{\eq}[1]{equation~(\ref{eq:#1})} %

\newcommand{\sect}[1]{section~\ref{sec:#1}} %

\newcommand{\cov}[2]{\mathrm{cov}\!\left({#1,\,#2}\right)} 

\DeclareSIUnit\electron{\ensuremath{\textnormal{e}^-}} 
\DeclareSIUnit\columns{\ensuremath{\textnormal{columns}}} 
\DeclareSIUnit\rows{\ensuremath{\textnormal{rows}}} 
\DeclareSIUnit\pixels{\ensuremath{\textnormal{pixels}}} 
\DeclareSIUnit\clight{\text{\ensuremath{\mathit{c}}}} 
\DeclareSIUnit\evperc{\eV/\clight} 

\DeclareMathOperator\erf{erf}

\makeatletter \newcommand*\fsize{\dimexpr\f@size pt\relax} \makeatother

\title{Timing measurements with a 3D silicon sensor on Timepix3 in a \SI{180}{\giga\evperc} hadron beam}

\author[a,1]{K.~Heijhoff,\note{Corresponding author.}}
\author[a]{K.~Akiba,}
\author[b]{R.~Bates,}
\author[a]{M.~van~Beuzekom,}
\author[a]{P.~Bosch,}
\author[a]{A.P.~Colijn,}
\author[a]{R.~Geertsema,}
\author[a]{V.~Gromov,}
\author[a]{M.L.E.~Heidotting,}
\author[a]{L.S.~Hendrikx,}
\author[a,2]{D.~Hynds,\note{Now at Department of Physics, University of Oxford, Oxford, United Kingdom}}
\author[a]{H.~Snoek}

\affiliation[a]{Nikhef, Amsterdam, the Netherlands}
\affiliation[b]{School of physics and astronomy, University of Glasgow, Glasgow, Scotland}

\emailAdd{k.heijhoff@nikhef.nl}

\abstract{Test beam measurements have been carried out with a 3D sensor on a Timepix3 ASIC and the time measurements are presented. The measurements are compared to those of a thin planar sensor on Timepix3. It is shown that for a perpendicularly incident beam the time resolution of both detectors is dominated by the Timepix3 front-end. The 3D detector is dominated by the time-to-digital conversion whereas the analog front-end jitter also gives a significant contribution for the thin planar detector. The 3D detector reaches an overall time resolution of \SI{567(6)}{\pico\second} compared to \SI{683(8)}{\pico\second} for the thin planar detector. For a grazing angle beam, however, the thin planar detector achieves a better time resolution because it has a lower pixel capacitance, and therefore suffers less from jitter in the analog front-end for the low charge signals that mainly occur in this type of measurement. Finally, it is also shown that the 3D and thin planar detector can achieve time resolutions for large clusters of about \SI{100}{\pico\second} and \SI{250}{\pico\second}, respectively, by combining many single hit measurements.}


\begin{document}
	\maketitle
	\flushbottom
	
	\section{Introduction}
	Future experiments at the High Luminosity LHC \cite{Apollinari:2017} will see a further increase in the number of concurrent events per bunch crossing leading to pile-up. A possible solution that will enable them to cope with the increased pile-up is 4D~tracking~\cite{Cartiglia:2017, Sadrozinski:2018} in which precise temporal information of the tracks helps the reconstruction algorithm to distinguish spatially overlapping vertices. Therefore, it is foreseen that precise time measurements will become crucial for vertex and tracking detectors used in particle physics experiments.
	
	New sensor technologies are being explored to achieve the time resolution required for 4D~tracking as conventional ``thick'' planar silicon pixel sensors provide inadequate resolution. One strategy to improve the time resolution of a sensor is to decrease the drift distance of charge carriers. This can be done, for example, by making thinner sensors \cite{Riegler:2017}. In doing so, however, the amount of signal charge is also reduced, which leads to an increase in jitter due to a decrease in signal-to-noise ratio. On the contrary, 3D silicon sensor technology \cite{Parker:1997} also reduces the charge carrier drift distance, but does not suffer from a reduction of the signal charge. However, the readout electrodes of these sensors typically have a larger capacitance, which also decreases the signal-to-noise ratio, but this time due to an increase in the noise instead. 
	
	In this paper timing measurements obtained with a 3D-silicon sensor bump bonded to a Timepix3 ASIC \cite{Poikela:2014} are presented, and compared to measurements obtained with a thin planar sensor also bonded to Timepix3. After a description of the sensors and the measurement setup, the time measurement mechanism of Timepix3 is discussed in detail. Then the results for particles crossing the sensors perpendicularly are discussed. After this, the results for particles at a grazing angle of incidence are presented, and finally the possibility of improving the time resolution by combining multiple hits on a track is explored. 
	
	\section{Experimental setup}
	\subsection{Description of sensors}
\begin{figure}[htbp]
		\centering
		\includegraphics{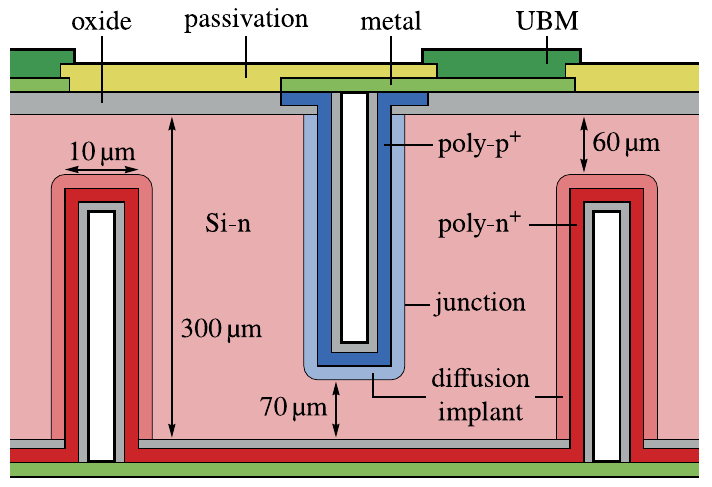}%
		\hskip 3mm%
		\includegraphics{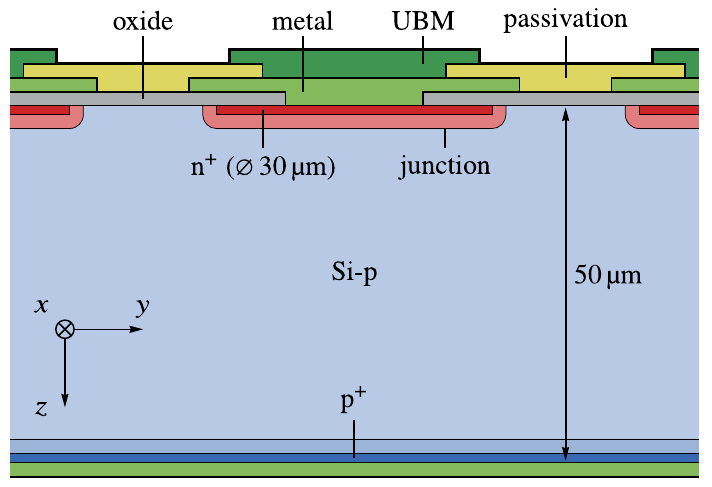}%
		\caption{Schematic diagrams of a double sided 3D (left) and a thin planar sensor (right). Dimensions and layout from measurements presented in this paper and \cite{AlipourTehrani:2017, Pellegrini:2008, MacRaighne:2011}. The local reference frame of the devices under test is also shown. The beam points along the negative $z$-direction for a perpendicular incidence.}
		\label{fig:sensorDiagrams}
	\end{figure}

	The 3D sensor technology differs from the planar technology by the geometry of the electrodes. This can be seen in the schematic diagrams of the sensors used in this study, which are shown in \fig{sensorDiagrams}. In a planar sensor, the pixel electrodes are made by implanting dopants at the bulk surface whereas in a 3D sensor the electrodes penetrate into the bulk. For both sensors, the backside is a single electrode where the bias potential is applied, and the frontside electrodes are connected to individual readout channels on the ASIC. When depleted, charge carriers in a planar sensor drift perpendicularly to the sensor surface under influence of the electric field. In a 3D sensor the charge carriers drift mostly parallel to the sensor surface towards (or away from) the electrodes that are connected to the readout channels. The 3D and thin planar sensors used in this study collect holes and electrons, respectively, at the readout electrodes, which are connected to the ASIC. The $\textnormal{n}^{+}$ electrode of the 3D sensor is referred to as the field electrode.

	The double-sided 3D sensor that is used in this study has been fabricated at IMB-CNM \cite{Pellegrini:2008}. The electrode regions were etched into the bulk material using an inductively coupled plasma. The high aspect ratio of the electrodes was achieved by a process of alternating etch and passivation cycles. The electrodes were then formed by filling the etched holes with doped polysilicon. This process was repeated for both the front- and backside of the sensor to make the $\textnormal{p}^{+}$ and $\textnormal{n}^{+}$ doped electrodes, respectively. Double sided processing of the wafer has multiple advantages over single sided processing: (i)~producing electrodes with different types of doping is more difficult on a single surface, (ii)~it makes it simpler to apply the bias potential, and (iii)~the electrodes don't penetrate through the whole sensor which means that there is still some active sensor material above the electrodes, which improves the efficiency \cite{MacRaighne:2011}. The double sided processing does, however, require an alignment step which increases the cost. The thin planar sensor used in this study was fabricated at Advacam \cite{Wu:2012} and is an active-edge sensor \cite{Nurnberg:2019} originally produced for the CLIC vertex detector \cite{Burrows:2018}. Both sensors are bonded to a Timepix3 ASIC that is read out by a SPIDR readout system~\cite{Visser:2015, Heijden:2017}.
	
	\subsection{Measurement setup}
\begin{figure}[htbp]
		\centering
		\includegraphics{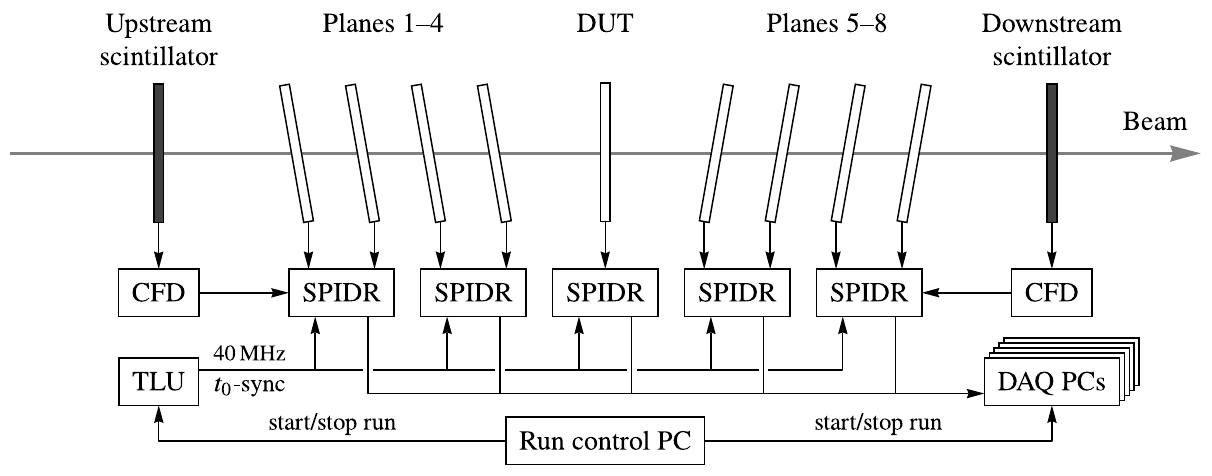}%
		\caption{Diagram of the LHCb VELO Timepix3 Telescope \cite{Heijhoff:2020}. The distance between the scintillators is about \SI{1}{\metre}, the distance between the outer telescope planes is about \SI{48}{\centi\metre}, and the distance between adjacent planes is about \SI{2.5}{\centi\metre}.}
		\label{fig:telescopeDiagram}
	\end{figure}

	The measurements for this study were performed at the H8 beam line of the CERN Super Proton Synchrotron~(SPS) using the LHCb VELO Timepix3 Telescope \cite{Akiba:2019,Heijhoff:2020}, which provides the track reconstruction by measuring the position of each particle on eight detector planes (\fig{telescopeDiagram}). Subsequently the track position is interpolated with a resolution of about \SI{1.6}{\micro\meter} to a device under test (DUT), which is located at the centre of the telescope. The DUT is mounted on two translation stages to align it with respect to the telescope planes, and a rotation stage to allow for angle studies. The particle beam consists of mixed hadrons ($p$, $\pi$, $K$) of about \SI{180}{\giga\evperc}. The hadrons are delivered in spills that are repeated every \SIrange[range-phrase=--, range-units=single]{20}{30}{\second} and contain a few million particles that are distributed over a duration of typically~\SI{4.5}{\second}. Two independent reference time measurements are provided for each particle by two fast scintillators with an active area of ${1.5\times\SI{1.5}{\square\centi\meter}}$. They are located up- and downstream of the telescope, and are equipped with constant fraction discriminators (CFD). Their signals are registered by the on\=/board time-to-digital converter (TDC) of the SPIDR readout system~\cite{Visser:2015, Heijden:2017}. The up- and downstream scintillators have time resolutions of respectively \SI{381(8)}{\pico\second} and \SI{182(4)}{\pico\second}~\cite{Heijhoff:2020}. 

	\subsection{Timepix3 calibration}
	\label{sec:tpx3Calibration}

\begin{figure}[htbp]
		\centering
		\includegraphics{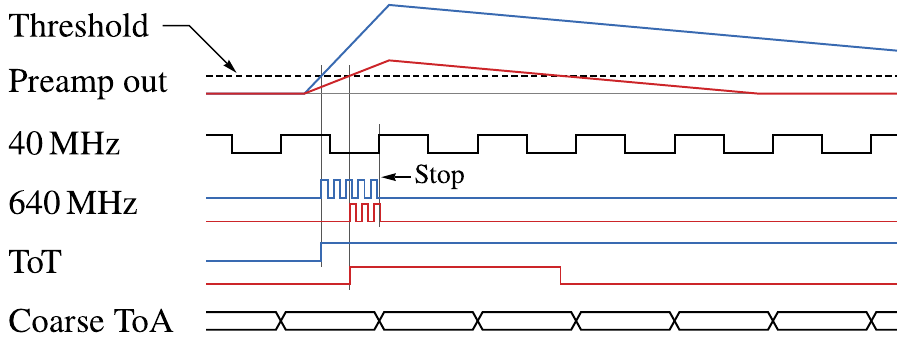}%
		\caption{Diagram of the ToA and ToT measurements in Timepix3 for two hits with a different signal amplitude \cite{Heijhoff:2020}.}
		\label{fig:tpx3Timing}
	\end{figure}

	For both DUTs the Timepix3 ASICs were operated in the ToA \& ToT acquisition mode in which both the time of arrival (ToA) and time over threshold (ToT) are measured for each hit. \Fig{tpx3Timing} illustrates how these measurements are performed in Timepix3. When the preamplifier output crosses a threshold value, a local voltage controlled oscillator (VCO) is started which has a frequency of \SI{640}{\mega\hertz}. The pixel logic determines the so-called fine time of arrival (fToA) by counting the number of clock cycles from the VCO until a rising edge of the \SI{40}{\mega\hertz} system clock arrives. Meanwhile, the pixel logic also registers the number of \SI{40}{\mega\hertz} clock cycles, which is called the coarse time of arrival (cToA). From the fToA and the cToA, the overall time of arrival of each hit can be determined with a granularity of about \SI{1.56}{\nano\second}. For the same hit, the time over threshold is determined with a granularity of \SI{25}{\nano\second} by counting the \SI{40}{\mega\hertz} clock while the preamplifier output is above the threshold value.
	
	When a minimum ionising particle (MIP) crosses the sensor, the generated electron-hole pairs induce a transient current signal on the pixel implants, which is subsequently integrated by the charge sensitive preamplifier in the analog front-end of the corresponding pixel \cite{Gaspari:2014}. The preamplifier output signal is proportional to the integrated current, and thus to the number of electron-hole pairs that were generated in the sensor. The integrated current is then discharged at a constant rate by the Krummenacher feedback of the preamplifier \cite{Krummenacher:1991}, and the output signal will therefore decay linearly. As a result, the ToT is roughly proportional to the amount of charge in the signal. To convert the ToT measurement to charge, a test-pulse calibration is performed. A controlled amount of charge can be injected into the analog front-end of each pixel using the built-in test-pulse circuitry. This is done for charge values up to about $\SI{18}{\kilo\electron}$ in steps of approximately $\SI{250}{\electron}$. The relationship between charge and the mean ToT response is then determined for each pixel by fitting the surrogate function \cite{Jakubek:2008}
	\begin{equation}
		\label{eq:chargeCalibration}
		\textnormal{ToT} = p_0 + p_1 Q - \frac{p_2}{Q - p_3}
		\,,
	\end{equation}
	and the inverse relationship gives the conversion from ToT value to signal charge.
	
	The VCO divides the \SI{25}{\nano\second} period of the \SI{40}{\mega\hertz} clock into \num{16} TDC time bins of approximately \SI{1.56}{\nano\second} each. However, there are significant deviations in the widths of these bins as a consequence of (i) variations in the VCO frequency due to process variation in the fabrication of Timepix3, and (ii) variations in the signal propagation delay between a pixel and its corresponding VCO (which is shared by eight pixels) due to differences in the capacitive loading of the traces that connect them. The latter affects the width of the first time bin, which has an fToA value of zero. The time bins for fToA values \numrange{1}{14} have a size that is mainly determined by the VCO frequency. The size of the last time bin, with $\textnormal{fToA}=15$, is determined by how much time in the \SI{25}{\nano\second} period remains after subtracting the total width of the other time bins. 
	
\begin{figure}[htbp]
		\centering
		\includegraphics{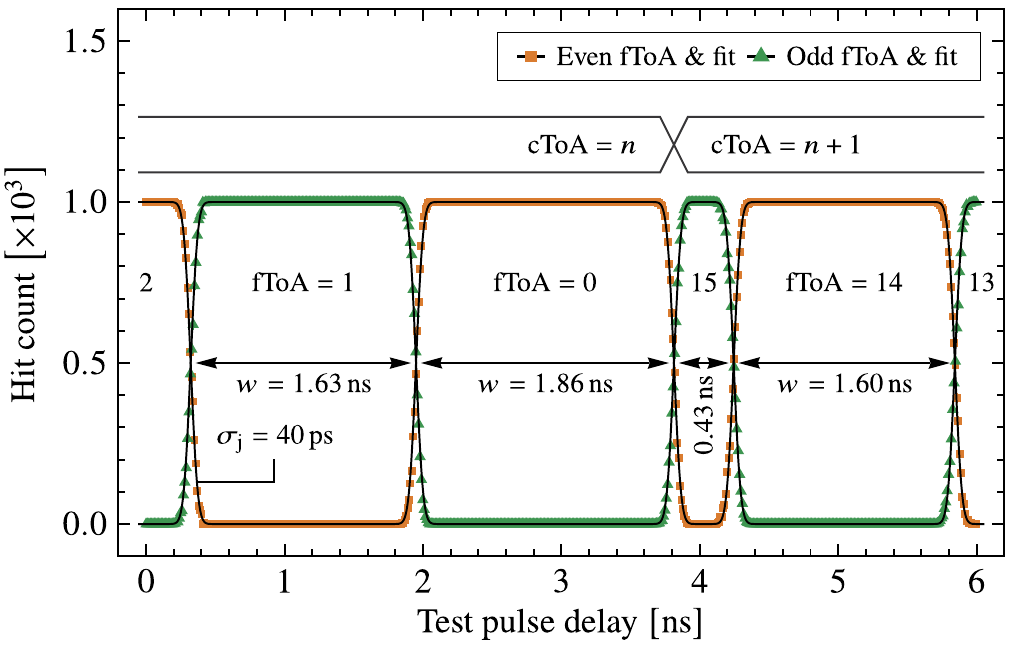}%
		\caption{Test pulse delay scan of a single pixel of the 3D detector. The plot shows the number of hits with even and odd fToA values as a function of the delay configuration of the pulse generator.}
		\label{fig:ds3dTpDelayScan}
	\end{figure}

	To correct timing errors introduced by the TDC time bin variations, their sizes are measured using externally timed digital test pulses, which bypass the analog front-end of the pixel, and directly go to the digital part instead. The external test pulses are generated by a pulse generator that is triggered on an edge of the \SI{40}{\mega\hertz} clock for synchronisation. The trigger delay is then varied in steps of \SI{10}{\pico\second} to scan the whole \SI{25}{\nano\second} period. For each value of the trigger delay, \num{1000} test pulses are sent to the pixels, and the resulting fToA values are recorded. \Fig{ds3dTpDelayScan} shows a part of such a delay scan for a single pixel of the 3D detector. For this pixel, a test pulse that is generated with a trigger delay of zero arrives in the $\textnormal{fToA}=2$ time bin\===a trigger delay of zero is not necessarily aligned with a \SI{40}{\mega\hertz} clock edge due to delays in the electronics and cabling. As the trigger delay is increased, the fToA decreases because the time to the first subsequent \SI{40}{\mega\hertz} edge decreases. After an fToA value of zero, the test pulse arrives in the $\textnormal{fToA}=15$ bin of the next \SI{25}{\nano\second} period. 
	
\begin{figure}[htbp]
		\centering
		\includegraphics{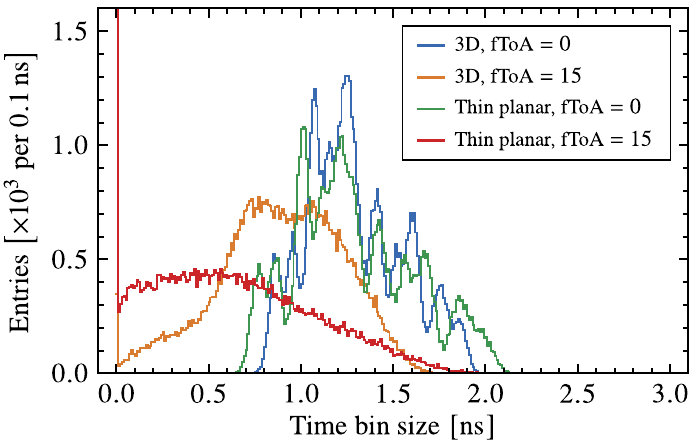}%
		\hskip5mm
		\includegraphics{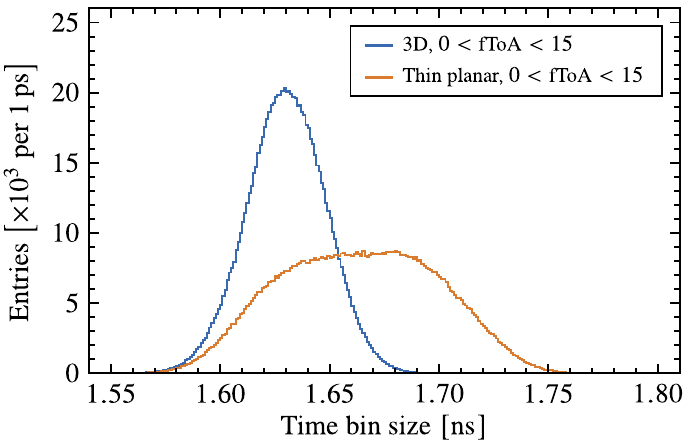}%
		\caption{Distribution of the TDC time bin size for the 3D and thin planar detector. The left plot shows the first and last bins which vary the most in size, and the right plot shows the time bins that lie in between. There is a spike at zero in the left plots due to bins with a size of zero.}
		\label{fig:timeBinWidths}
	\end{figure}

	For each fToA value $n$, the time bin size is determined by fitting the number of hits in its corresponding time bin with
	\begin{equation}
		\label{eq:hitCountVsTpDelay}
		N_n\!\left(d\right) = 
		\frac{N_{\textnormal{total}}}{2} 
		\left[
			\erf\!\left(\frac{d - l}{\sqrt{2}\,\sigma_{\textnormal{j}}}\right)
			- \erf\!\left(\frac{d - l - w}{\sqrt{2}\,\sigma_{\textnormal{j}}}\right)
		\right]\, ,
	\end{equation}
	where $N_{\textnormal{total}}$ is the total number of test pulses per delay value, $d$~is the trigger delay, $\sigma_{\textnormal{j}}$ is the jitter in the measured arrival time of the test pulses with respect to the \SI{40}{\mega\hertz} clock, and finally, $l$ and $w$ are the lower edge and size of the time bin, respectively. \Fig{timeBinWidths} shows the distribution of the time bin size for both DUTs. It can be seen that the first and last time bins deviate the most from their design value of \SI{1.56}{\nano\second}. The results also show that some pixels only have 15 non-zero time bins: \SI{0.49}{\percent} and \SI{24.8}{\percent} for the 3D and thin planar detector, respectively. The difference between the two detectors is explained by the fact that they use different versions of Timepix3. The 3D detector uses the first iteration of Timepix3 whereas the thin planar detector uses the second iteration in which a power distribution issue was addressed. The results for the 3D detector are in agreement with test pulse delay scans performed on other first-iteration Timepix3 chips \cite{Zappon:2015}. For each hit, the time difference between the rising edge of the \SI{40}{\mega\hertz} clock and the centre location of the time bin for the corresponding fToA value is subtracted from the cToA to correct for the unequal sizes of the TDC time bins.

	As was mentioned above, each VCO is shared by a group of eight pixels called a superpixel. This can lead to the scenario in which the oscillator is already running when a hit arrives in a pixel. The arrival time of the earlier hit that started the oscillator, lies somewhere in a \SI{1.56}{\nano\second} range (assuming an ideal VCO). The exact arrival time in this range is (of course) unknown, and the location of the time bin of the second hit is therefore also unknown because it depends on when the oscillator was started. As a consequence, the time bins will have a difference in time resolution: For the first hit, the time binning contribution $\sigma_{\textsc{tdc}}$ to the overall time resolution is $\SI{1.56}{\nano\second}/\sqrt{12}$ because the first time measurement is described by a rectangular distribution. The time measurement of the second hit, however, has a (symmetric) triangular distribution with a base of $2\times\SI{1.56}{\nano\second}$ due to the unknown arrival time of the first hit, and therefore it has a time binning resolution of $2\times\SI{1.56}{\nano\second}/\sqrt{24}$, which is a factor $\sqrt{2}$ worse than that of the first hit. Since the second hit is typically also associated with more timewalk, and therefore also more timing jitter (due to a lower signal to noise ratio in the analog front-end), only the time measurements of the first hits in each superpixel are used in this study.

\begin{figure}[htbp]
		\centering
		\includegraphics{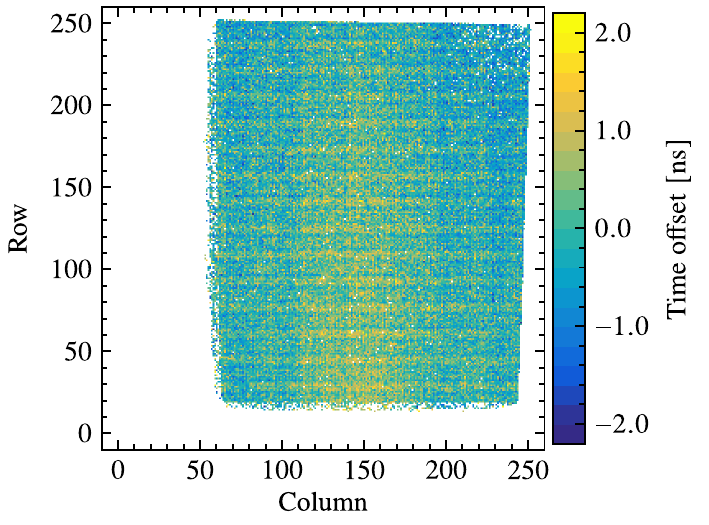}%
		\hskip3mm
		\includegraphics{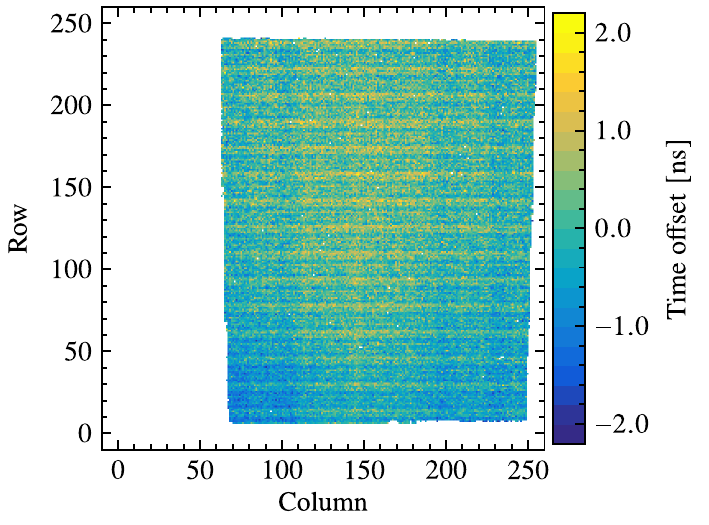}%
		\caption{Pixel time offsets of the 3D (left) and thin planar detector (right) after correcting the timing errors from VCO variations.}
		\label{fig:thplPixelOffsets}
	\end{figure}

	For each pixel, the time of arrival within the \SI{25}{\nano\second} period is determined as the centre location of the TDC time bin with respect to the \SI{40}{\mega\hertz} clock. However, this time calibration only corrects for timing errors within a single clock phase. A pixel can still have an overall time offset due to (i)~phase differences among pixels in the \SI{40}{\mega\hertz} clock due to the clock distribution, and (ii)~variations in the speed of the analog front-end due to the power distribution over the pixel matrix. These offsets cannot be measured with test pulses because they themselves suffer from (unknown) differences in arrival time over the pixels due to their routing delays in the chip. Test beam data is used to determine these overall time offsets as the mean time-residual with respect to the downstream scintillator for each pixel. These values are then used as corrections and subtracted from all time measurements in those pixels. \Fig{thplPixelOffsets} shows the pixel time offsets for both DUTs. The pixel time offsets of the 3D and thin planar detectors approximately follow Gaussian distributions with standard deviations of \SI{0.64}{\nano\second} and \SI{0.56}{\nano\second}, respectively. The difference in their global behaviour over the pixel matrix is attributed to the fact that they use different versions of Timepix3.

	\section{Measurements and results}
	\subsection{Perpendicular incidence}
	\label{sec:perpendicularBeam}
	In this section the timing performance of the DUTs is assessed with a perpendicularly incident beam with particles crossing the sensor in the negative $z$-direction (see \fig{sensorDiagrams} for reference). To achieve a perpendicular incidence, the mean cluster size was measured for various angular positions of the rotation stage ranging from \SI{-8}{\degree} to \SI{8}{\degree}, and the angular position corresponding to a minimum in the mean cluster size, and thus true perpendicular incidence, was determined by fitting a second degree polynomial.
	
	The time resolution of a sensor typically improves with an increase in charge carrier velocity; therefore, the focus in this section is mainly on measurements performed at the highest reverse bias potential that allowed operation of the detector without breakdown---a state in which the leakage current increases exponentially. The 3D and thin planar detectors were operated at \SI{60}{\volt} and \SI{90}{\volt}, respectively. The threshold values for detecting a hit were set at \SI{800}{\electron} and \SI{700}{\electron}, respectively. First, the timing behaviour within a pixel cell for both DUTs is discussed. This is then shown to strongly depend on the typical signal size, which is substantially different for both sensors. Then the efficacy of two different types of timewalk corrections that can be applied to the time measurements is discussed. Finally the hit time resolution of both DUTs will be presented.
	
\begin{figure}[htbp]
		\centering
		\includegraphics{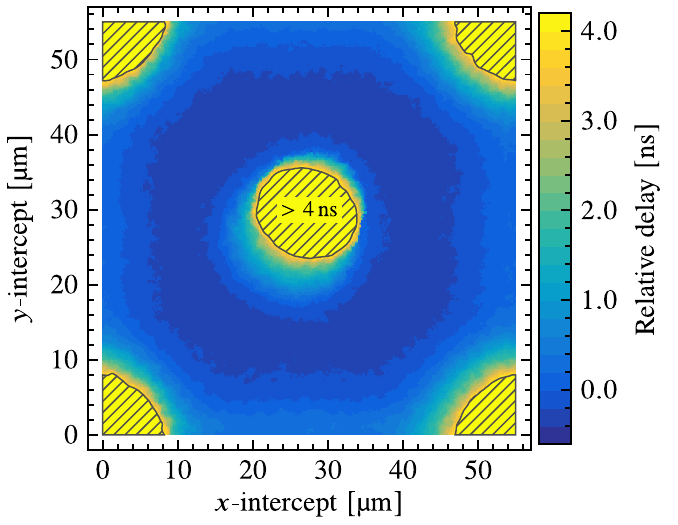}%
		\hskip5mm
		\includegraphics{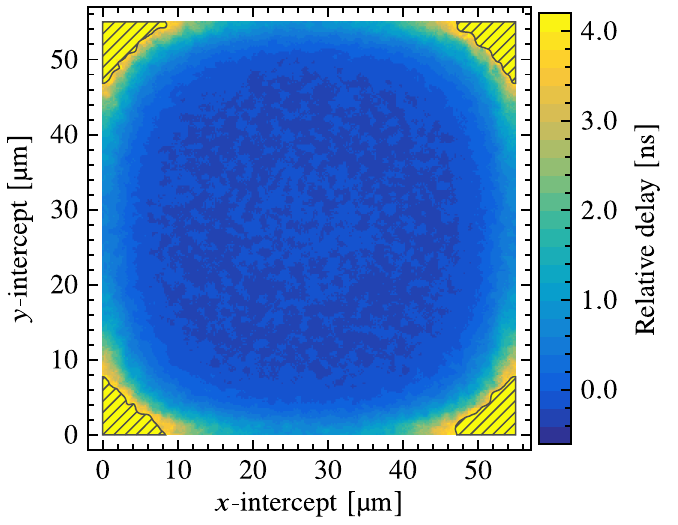}%
		\caption{Relative delay within a pixel cell of the 3D~detector~(left) and the thin planar detector~(right). The shaded regions indicate where the relative delay is longer than \SI{4}{\nano\second}.}
		\label{fig:intrapixelDelay}
	\end{figure}

	The track information provided by the telescope is used to the determine the track intercept with the DUT for each track. Events are then collected based on the intrapixel coordinates of the track intercept into (overlapping) circular bins of \SI{1}{\square\micro\meter} that are placed on a \SI{0.2}{\micro\meter} square grid. This spacing is significantly smaller than the \SI{1.6}{\micro\meter} resolution of the track intercept, and is chosen so to study the transition between the electrode regions and the region between them. For each bin the relative delay is determined as the mean difference between the hit time measurements on the Timepix3 ASIC and their corresponding reference time, which are defined as the weighted means of the up- and downstream scintillator measurements. It should be noted here that the overall time offset between the hits and the scintillators has been subtracted, and that a relative delay of zero therefore corresponds to this overall offset. The result is shown in \fig{intrapixelDelay} for both DUTs. The electrodes in the 3D sensor are clearly visible as regions that have a large positive delay. Furthermore, the readout electrode appears to be slightly off-centre and the time delay seems to increase more gradually on its bottom left side. For the thin planar sensor it can be seen that it is mainly slower near the pixel corners. This will be explained in terms of signal charge below. 

\begin{figure}[htbp]
		\centering
		\includegraphics{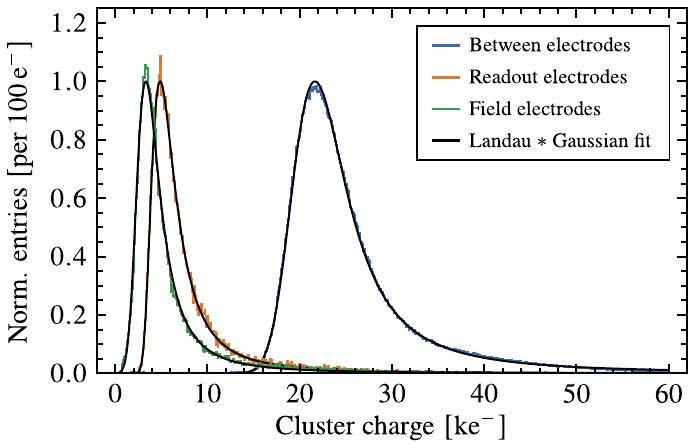}%
		\hskip5mm
		\includegraphics{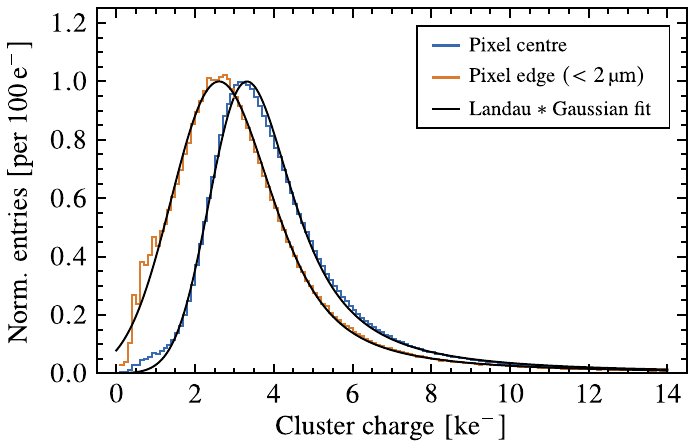}%
		\caption{Distributions of the collected charge per cluster for the 3D detector~(left) and the thin planar detector~(right) normalised to the peak values of their corresponding fits.}
		\label{fig:chargeDistributions}
	\end{figure}

	\Fig{chargeDistributions} shows the cluster charge distributions obtained in these measurements for both DUTs together with fits of a Landau distribution convolved with a Gaussian distribution. The left plot shows the charge distribution for particles going through different region of the 3D sensor. Events from particles going through an electrode have less charge because only energy that is deposited in the bulk silicon is converted into electron-hole pairs that drift so that they induce a signal. The cluster charge resulting from a particle going between the electrodes has a most probable value (MPV) of about \SI{22}{\kilo\electron}, which is in agreement with a MIP crossing \SI{300}{\micro\meter} of silicon. The cluster charge in the thin planar sensor has an MPV of about \SI{3.3}{\kilo\electron} for particles going through the central area of a pixel (defined as the region where the track intercept is at least \SI{2}{\micro\meter} away from the nearest pixel edge). This value is as can be expected for a MIP crossing \SI{50}{\micro\meter} of silicon. It can also be seen that the MPV of the cluster charge from particles crossing the sensor close to the boundary between two pixels is lower. This is expected because charge is being shared by two (or more) pixels, and sometimes not all pixels collect enough charge to reach the threshold level for registering a hit, causing this charge to escape detection and resulting in a lower cluster charge measurement. Furthermore, it seems that the thin planar detector has hits that are below threshold, but this is probably a problem with the charge calibration (\sect{tpx3Calibration}) for small charges due to the ToT distribution being partially cut off, which leads to a mean ToT value that is not representative for measurements of charges close to the threshold value, which in turn affects the fit of the surrogate function.
	
\begin{figure}[htbp]
		\centering%
		\lineskip=0pt%
		\includegraphics{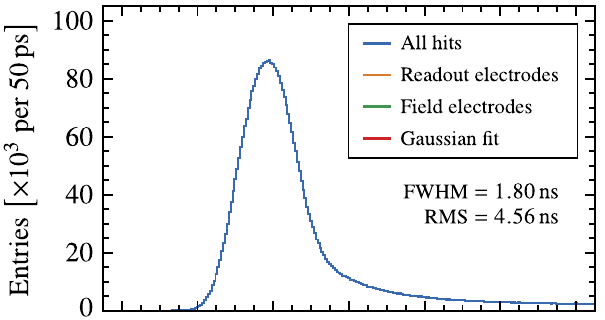}%
		\hskip2mm%
		\includegraphics{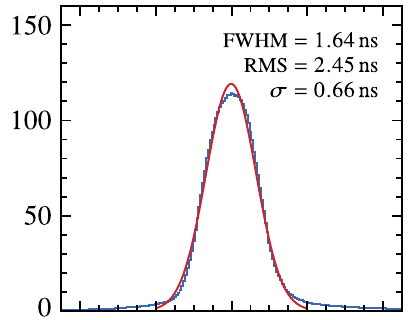}%
		\includegraphics{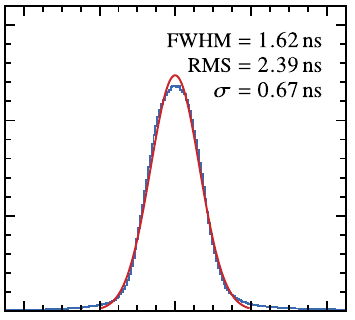}\\*[0pt]%
		\includegraphics{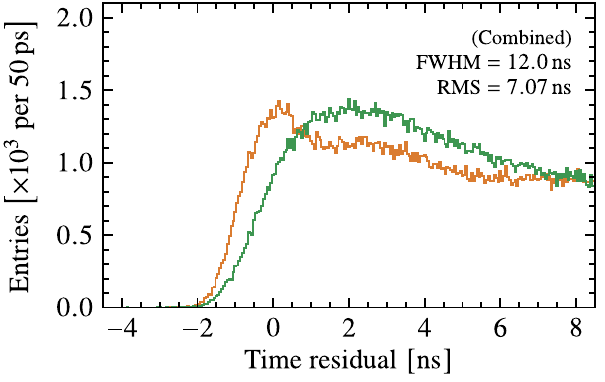}%
		\hskip2mm%
		\includegraphics{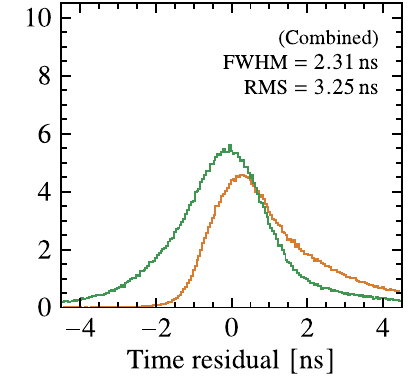}%
		\includegraphics{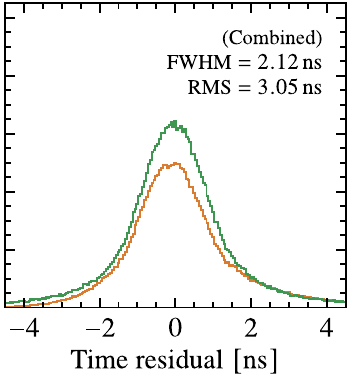}%
		\caption{Time residuals in three different regions of the 3D detector without timewalk correction (left), with a timewalk correction based on charge only (centre), and with a correction based on both charge as well as track intercept (right). This shows that the tail in the time residual distribution can be mostly corrected with only a charge-based timewalk correction.}%
		\label{fig:ds3dTimeResiduals}%
	\end{figure}

	\Fig{ds3dTimeResiduals} shows the time residuals for the 3D~detector after applying various corrections. In the left plot the time measurements are only corrected for the systematic timing offsets in the pixel matrix as described above in \sect{tpx3Calibration}. It can be seen that there is a tail in the time residual distribution that is dominated by hits from particles going through one of the electrodes. The middle plot shows the same time residuals after they are also corrected for timewalk by first collecting hits into bins based on their charge measurement, and subsequently subtracting the mean of the time residuals in each bin. This correction significantly narrows the readout and field electrode distributions because these measurements suffer more from timewalk due to the lower signal charge as was shown in \fig{chargeDistributions}. It can also be seen that the two residual distributions of the electrode events are not aligned after the timewalk correction. Somewhat surprisingly, the readout electrode region appears to be slower. As will be shown below, there is a slow region above the readout electrodes that can explain why these events are late. The right plot shows the result of a method that also corrects time variations that are not due to signal size variations. This method works by also binning hits on the track intercept within the pixel (in addition to binning on charge) leading to a lookup table of corrections in terms of the $x$-intercept, $y$-intercept, and charge. This effectively corrects for spatial regions that are slower (or faster) than others. In the remainder of this paper these two types of corrections will be referred to as partial- and full timewalk~corrections, respectively. The term ``timewalk'' is usually restricted to only describe those variations in time-to-threshold (the time it takes a signal to reach threshold value) that are due to variations in signal size, but for conciseness this definition is expanded to include also other systematic effects that affect the time-to-threshold, such as variations in signal induction affecting the signal shape (which also includes drift time effects). Although applying a full timewalk correction can be important in some cases \cite{Heijhoff:2020}, its impact on the time resolution is relatively small for these DUTs as will be shown shortly.
	
\begin{figure}[htbp]
		\centering
		\lineskip=0pt%
		\includegraphics{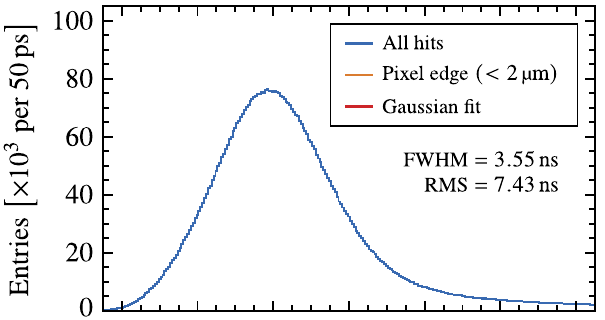}%
		\hskip2mm%
		\includegraphics{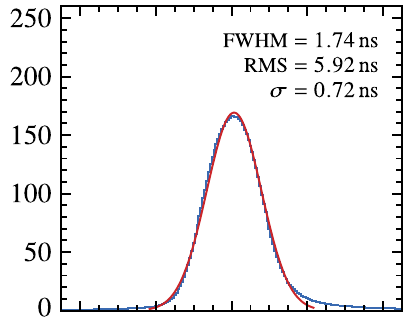}%
		\includegraphics{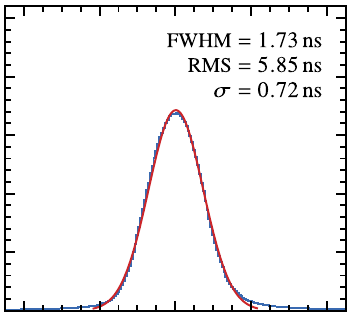}\\*[0pt]%
		\includegraphics{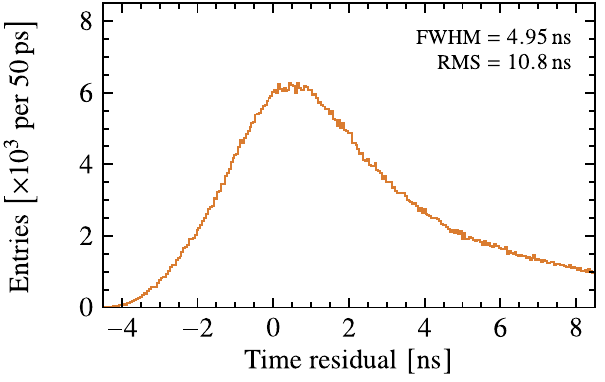}%
		\hskip2mm%
		\includegraphics{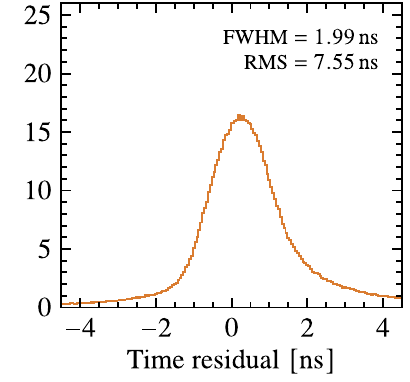}%
		\includegraphics{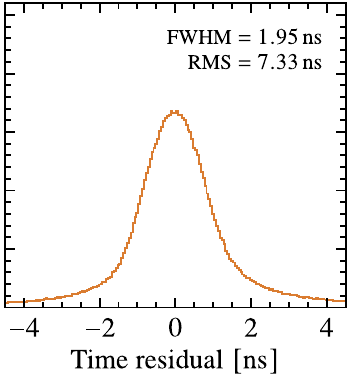}%
		\caption{Time residuals in two different regions of the thin planar detector without timewalk correction (left), with a timewalk correction based on charge only (centre), and with a correction based on both charge as well as track intercept (right). The timewalk corrections improve the residuals significantly because the sensor has low charge signals.}%
		\label{fig:thplTimeResiduals}%
	\end{figure}

	\Fig{thplTimeResiduals} shows the time residuals for the thin planar detector. The left plot, which contains the time residuals without timewalk correction, shows that the distribution is wider than that of the 3D detector. This is due to more severe timewalk effects because the signals in the thin planar sensor are typically much smaller (see \fig{chargeDistributions}). Also, the effect is more pronounced for particles that cross the sensor close to the edge of a pixel due to charge sharing. The middle plot shows that the partial timewalk correction, which is based on charge only, narrows the distribution considerably. The distribution from events near the pixel edge is still slightly off-centre, which is most likely due to a slower signal induction as a consequence of the electric field shape as well as a nonuniform weighting field. This is caused by the relatively small size of the pixel implant, which has a diameter of \SI{30}{\micro\meter} compared to the pixel pitch of \SI{55}{\micro\meter}. Applying a full timewalk correction corrects for the remaining time offsets.
	
	The resolution of the hit time measurement is determined using the two residuals with respect to the up- and downstream scintillators. Since the scintillators perform independent measurements, the covariance between the two residuals corresponds to the time resolution of the hit time measurement: 
	\begin{equation}
		\label{eq:timeResolution}
		\sigma_{t}^2=\cov{t - t_{\textnormal{d}}}{t - t_{\textnormal{u}}}
	\end{equation}
	where $t$ is the hit time, and $t_{\textnormal{u}}$ and $t_{\textnormal{d}}$ are the up- and downstream scintillator measurements \cite{Heijhoff:2020}. The covariance is determined by performing a maximum likelihood fit of a bivariate Gaussian distribution to the residuals. 
	
\begin{figure}[htbp]
		\centering
		\includegraphics{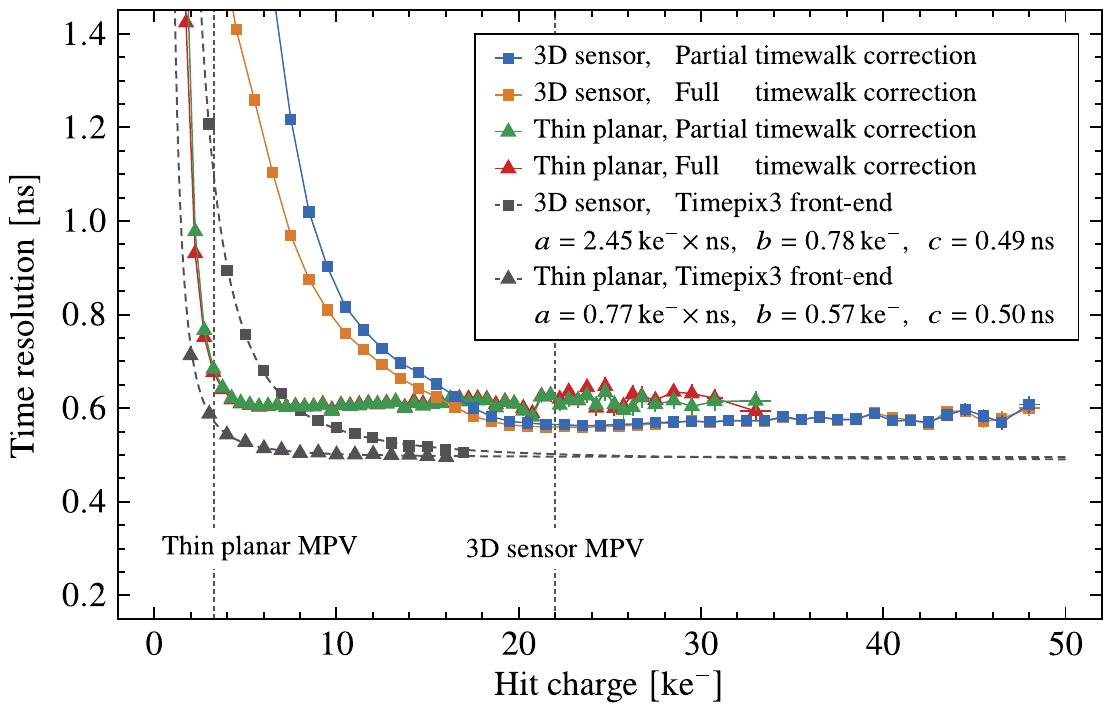}%
		\caption{The time resolution as a function of hit charge for both DUTs. The contribution of the Timepix3 front-end obtained from test pulses is also shown.}
		\label{fig:timeResolution}
	\end{figure}

	\Fig{timeResolution} shows the hit time resolution of both DUTs as a function of signal charge. It also shows the combined contribution to the overall time resolution of the analog and digital parts of the front end in Timepix3. This contribution represents the limit of what time resolution can be achieved with Timepix3 and depends on the pixel capacitance of the sensor. A measurement of the front-end contribution is briefly explained in the following paragraph before discussing the test beam results shown in the figure.

	The Timepix3 analog front-end time resolution was measured using the same method as the test pulse delay scan that was used to determine the sizes of the time bins in \sect{tpx3Calibration}, but instead of sending the test pulses directly to the digital front-end, they were sent to the analog front-end. The delay scan was repeated for test pulse amplitudes corresponding to injected charges $Q$ ranging from \SI{2}{\kilo\electron} up to \SI{17}{\kilo\electron} in steps of \SI{1}{\kilo\electron}. The maximum amount of injected charge is limited by the internal DACs that provide the test pulse voltage. The parameter $\sigma_{\textnormal{j}}$ in \eq{hitCountVsTpDelay} is now identified as the jitter contribution of the analog front-end:
	\begin{equation}
		\sigma_{\textnormal{j}} = \frac{\sigma_{\textnormal{v}}}{dV\!/dt}
		\, ,
	\end{equation}
	where $\sigma_{\textnormal{v}}$ and $dV\!/dt$ are the noise and slew rate of the preamplifier output signal at the threshold value. The combined time resolution of the front-end is obtained by adding the contribution of the digital front-end to the fit result as
	\begin{equation}
		\sigma_{\textnormal{fe}}^2 = \left(\frac{\sigma_{\textnormal{v}}}{dV\!/dt}\right)^2 + \frac{1}{n}\sum_{i=1}^n \frac{w_i^2}{12}
		\, ,
	\end{equation}
	where the sum is over the time bin sizes $w_i$ that were measured before. The second term thus describes the mean variance of rectangular distributions having widths $w_i$, and taking its square root gives \SI{461}{\pico\second} and \SI{473}{\pico\second} for the 3D and thin planar detectors, respectively. The thin planar detector has a higher value due to its bigger time bins (see also \fig{timeBinWidths}). The charge dependence of the front-end time resolution is modelled as
	\begin{equation}
		\sigma_{\textnormal{fe}}\!\left(Q\right)^2 = \left(\frac{a}{Q-b}\right)^2 + c^2
		\, ,
	\end{equation}
	where $a$, $b$, and $c$ are fit parameters. The fits are shown for the front-ends of both DUTs. It can be seen that the thin planar detector has a lower analog front-end contribution to the time resolution because the sensor has a lower pixel capacitance than the 3D sensor.

	\Fig{timeResolution} shows that the hit time resolutions of both DUTs have a strong charge dependence. The time resolution of the 3D detector after partial and full timewalk correction is dominated by the analog front-end for signals that are larger than \SI{10}{\kilo\electron} and \SI{11.5}{\kilo\electron}, respectively. For the thin planar detector the time resolution is dominated by the analog front-end for signals larger than \SI{2}{\kilo\electron}. After partial and full timewalk corrections, the 3D detector achieves an overall resolution of \SI{620(7)}{\pico\second} and \SI{609(7)}{\pico\second}, respectively. This is only marginally better than the standard \SI{300}{\micro\meter} planar sensors of the telescope which achieve a resolution of \SI{650(9)}{\pico\second} \cite{Heijhoff:2020}. When a minimum charge cut of \SI{15}{\kilo\electron} is applied, effectively rejecting events in the electrode regions, these figures improve to \SI{573(6)}{\pico\second} and \SI{567(6)}{\pico\second}, respectively. However, this cut also reduces the efficiency to \SI{75.2(15)}{\percent}. For the thin planar detector a minimum charge cut of \SI{1}{\kilo\electron} is used. It achieves an overall time resolution of \SI{683(8)}{\pico\second} after full timewalk correction. This time resolution is different than the time resolution of \SI{0.86}{\nano\second} found in \cite{Pitters:2019} for the same sensor type and ASIC. The difference can be attributed to several factors: (i) in addition to pixel corrections, fToA corrections were applied as described in \sect{tpx3Calibration}; (ii) the measurements in this study were performed at a higher bias potential; and (iii) in this study the timewalk correction has been determined using test beam data instead of test pulse measurements.
	
	\subsection{Grazing incidence}
\begin{figure}[htbp]
		\centering
		\includegraphics{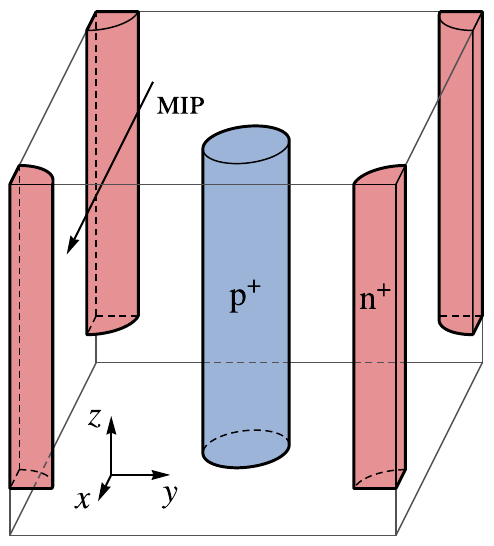}%
		\hskip20mm
		\includegraphics{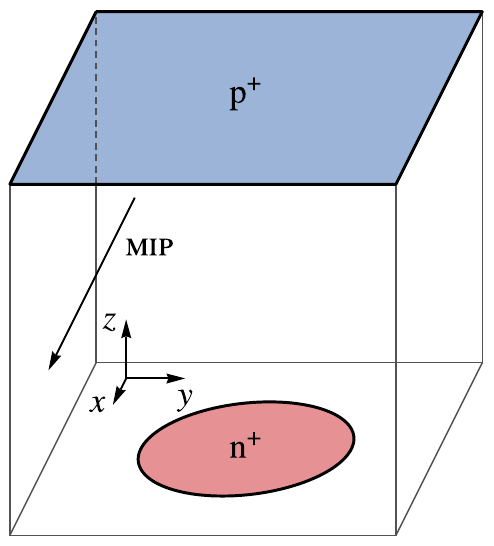}%
		\caption{Diagrams showing the beam direction in a pixel cell of the 3D detector~(left) and the thin planar detector~(right) for measurements performed with a grazing-incidence beam.}
		\label{fig:grazingDiagrams}
	\end{figure}

	For the measurements discussed in this section, the DUTs are rotated around their local $y$-axes such that the beam points mostly in the $x$-direction with a small positive $z$-component (see \fig{grazingDiagrams}). This results in long clusters covering \num{192.0(11)} and \SI{145.7(24)}{\columns} for the 3D and thin planar sensor, respectively. This allows for the investigation of the timing behaviour at various depths in the sensor as particles traverse a $z$-range of less then \SI{2}{\micro\meter} in each pixel. A more detailed explanation of this method can be found in \cite{dallocco:2021}. First the charge collection as well as the relative delay within a pixel cell of both DUTs is discussed, after which the single hit time resolution is assessed. The final part of this section investigates the possibility of combining the hit time measurements of each cluster to obtain a more precise time measurement.

\begin{figure}[htbp]
		\centering
		\includegraphics{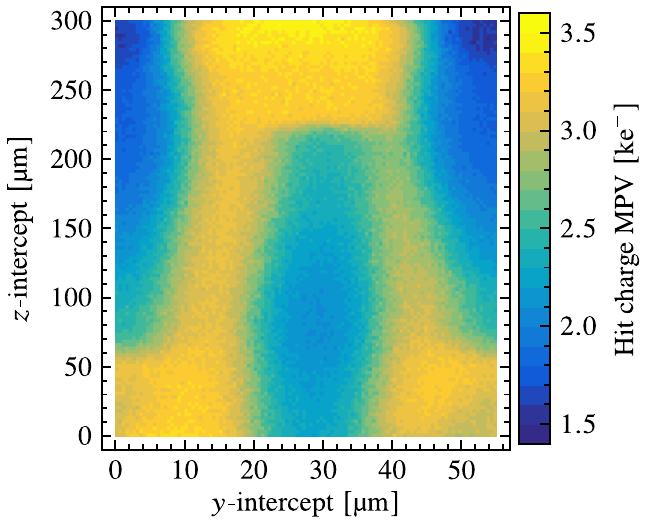}%
		\hskip5mm
		\includegraphics{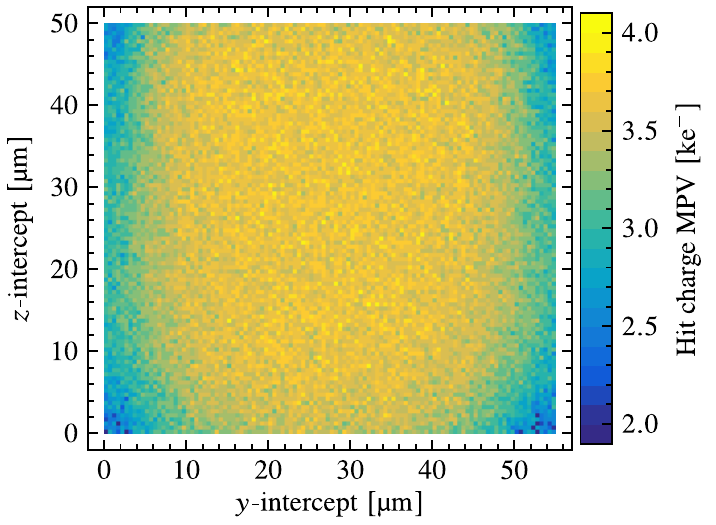}%
		\caption{Lateral intrapixel MPV of the collected charge for single hits of the 3D detector (left) and the thin planar detector (right) measured with a grazing incidence beam at bias potentials of \num{60} and \SI{90}{\volt}, respectively. The length of the pillar structures can be seen clearly from the abrupt changes in the charge distribution at about $z=\SI{230}{\micro\metre}$ and \SI{60}{\micro\metre} for the readout- and field pillars, respectively.}
		\label{fig:intrapixelChargeMpvGrazing}
	\end{figure}

	Hits are collected into spatial bins based on the $y$-, and $z$-intercepts of the reconstructed track with the $yz$-plane at the $x$-centre of the corresponding pixel. They are collected into rectangular bins of $0.5\times\SI{2.5}{\square\micro\meter}$ for the 3D detector, and square bins of $0.5\times\SI{0.5}{\square\micro\meter}$ for the thin planar detector. For each spatial bin, a convolution of a Landau and a Gaussian distribution is fitted to the charge distribution by performing a $\chi^2$ minimisation. \Fig{intrapixelChargeMpvGrazing} shows the MPV of the charge distribution as a function of $y$- and $z$-intercept. Compared to the measurements that were performed at perpendicular incidence, the collected charge in a single hit of the 3D detector is now much smaller because each pixel now only collects the charge from a particle traversing the width of the pixel instead of the full sensor thickness. As before, particles crossing the electrode regions of the 3D sensor have smaller signals. It can also be seen that the readout electrodes are under a slight angle, which explains the behaviour of the relative delay near the readout electrode in \fig{intrapixelDelay}. The signal size of the thin planar sensor is mostly uniform over the pixel except near the edges where charge is lost due to charge sharing.
	
\begin{figure}[htbp]
		\centering
		\includegraphics{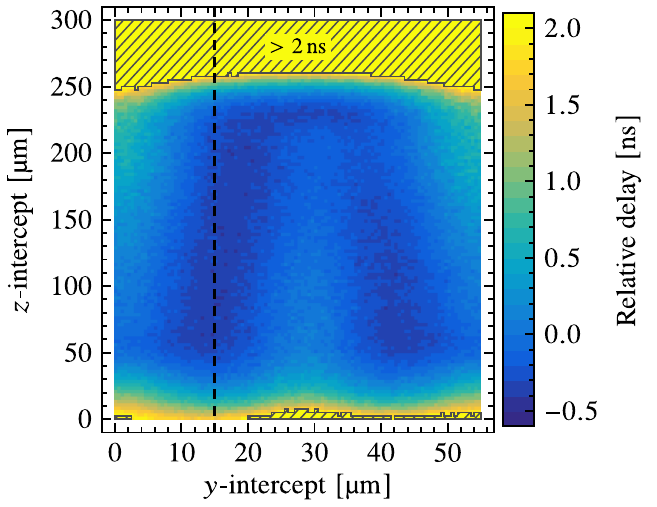}%
		\includegraphics{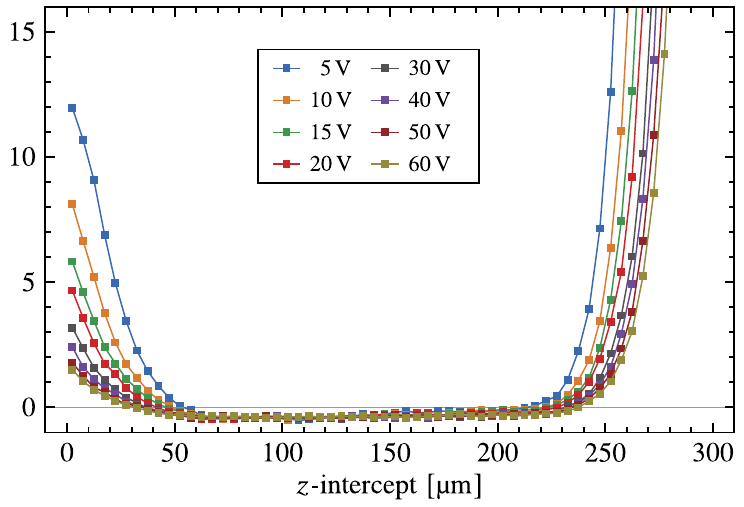}%
		\caption{Relative delay in the 3D detector measured at a bias potential of~\SI{60}{\volt} (left) and the relative delay at $y=\SI{15}{\micro\metre}$ as a function of sensor depth $z$ for various bias potentials (right). In both plots a partial timewalk correction is applied. The shaded region indicates where the relative delay is longer than~\SI{2}{\nano\second}.}
		\label{fig:intrapixelDelayGrazingDs3d}
	\end{figure}

\begin{figure}[htbp]
		\centering
		\includegraphics{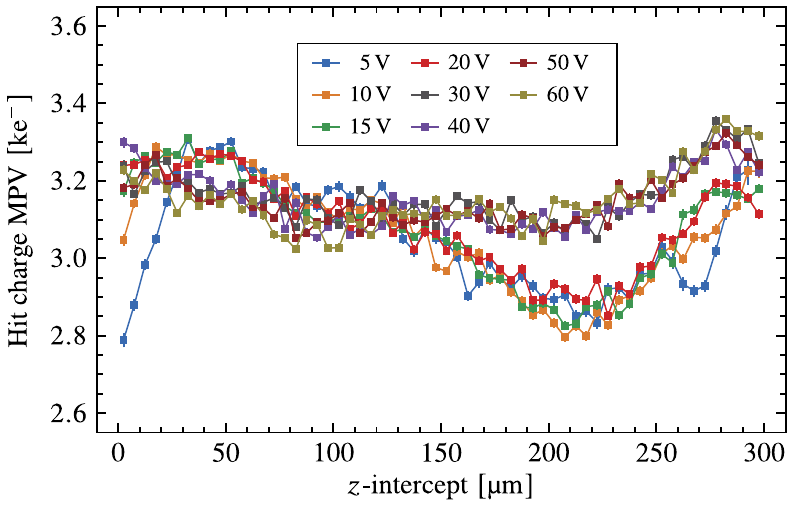}
		\caption{Hit charge MPV of the 3D detector as a function of depth for various bias potentials at $y=\SI{15}{\micro\meter}$.}
		\label{fig:ds3dChargeMpvBiasScanGrazing}
	\end{figure}

	The relative delay in each spatial bin is determined as before in \sect{perpendicularBeam}. \Fig{intrapixelDelayGrazingDs3d} shows the relative delay for the 3D detector after a partial timewalk correction was applied. It also shows a bias potential scan of the relative delay as a function of sensor depth $z$ for events that fall into a \SI{0.2}{\micro\meter} window centred at $y=\SI{15}{\micro\meter}$. Most notable is the  region above above $z\sim\SI{250}{\micro\meter}$ where the relative delay keeps increasing with $z$ until hits start falling outside of the $\SI{250}{\nano\second}$ time window of the clustering algorithm in the telescope reconstruction software. This is most likely an indication that the sensor is not depleted in this region, and that the charge is therefore being collected only by diffusion, resulting in long charge collection times. It might be expected that these long collection times allow for charge carrier recombination, but \fig{ds3dChargeMpvBiasScanGrazing} shows that there is no significant decrease in the collected charge in the nondepleted region. The carrier lifetime in lowly doped silicon (less than $\sim\SI[retain-unity-mantissa = false]{1e16}{\per\cubic\centi\meter}$) is dominated by the Shockley-Read-Hall mechanism \cite{ShockleyRead:1952,Hall:1952} in which electron-hole pairs recombine through deep-level impurities \cite{Schroder:1997}. This mechanism depends on the number of impurities and crystal defects in the bulk silicon. The actual carrier lifetime is therefore difficult to predict, but can be in the order of milliseconds for good quality silicon \cite{Schroder:2006}.
	
\begin{figure}[htbp]
		\centering
		\includegraphics{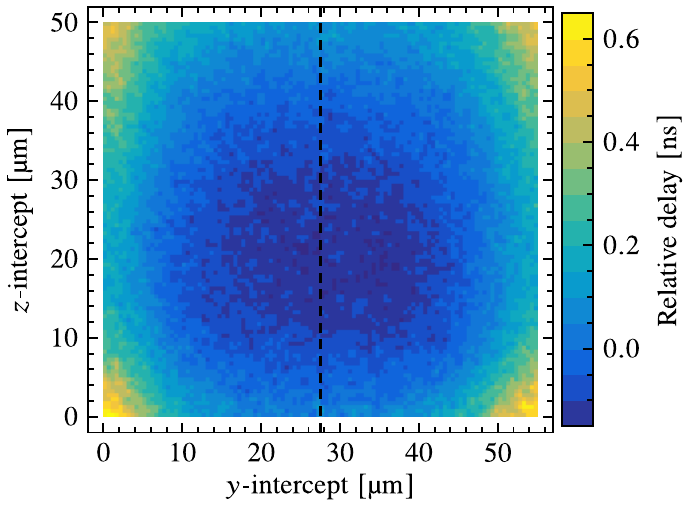}%
		\includegraphics{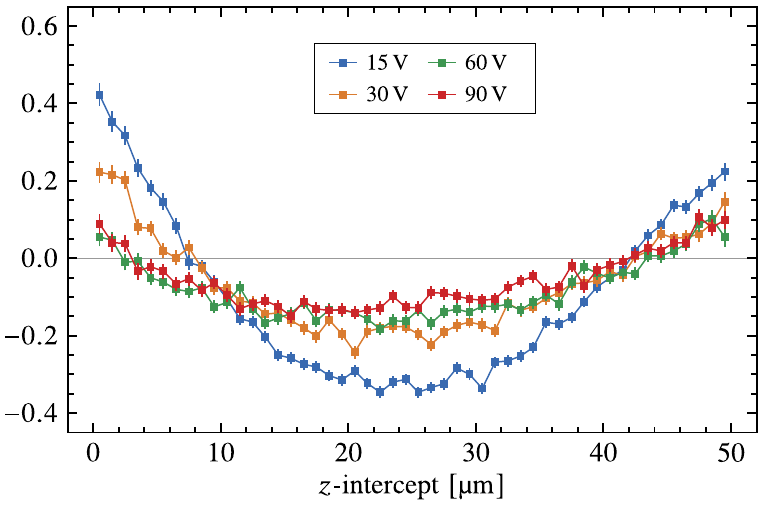}%
		\caption{Relative delay in the thin planar detector measured at a bias potential of~\SI{90}{\volt} (left) and the relative delay at $y=\SI{27.5}{\micro\metre}$ as a function of sensor depth $z$ for various bias potentials (right). In both plots a partial timewalk correction is applied.}
		\label{fig:intrapixelDelayGrazingThpl}
	\end{figure}

\begin{figure}[htbp]
		\centering
		\includegraphics{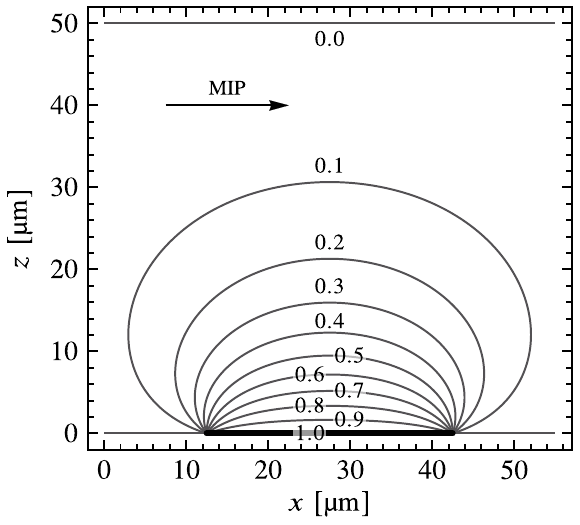}%
		\caption{Equipotentials of the analytic weighting potential (as derived in \cite{Riegler:2014}) at $y=\SI{27.5}{\micro\meter}$ for a \SI{30}{\micro\meter} square pixel implant.}
		\label{fig:thplWeightingPotential}
	\end{figure}

	The relative delay for the thin planar detector after a partial timewalk correction is shown in \fig{intrapixelDelayGrazingThpl} together with a bias potential scan of the relative delay as a function of sensor depth $z$ at $y=\SI{27.5}{\micro\meter}$. There is an increase in the relative delay near the $y$-edges of the pixel due to a slower signal induction. It can also be seen that the relative delay has a minimum near the $z$-centre of the sensor. This minimum is located at $z\sim\SI{20}{\micro\meter}$ for a bias potential of \SI{90}{\volt}. The fact that the sensor is faster in this region cannot be caused by variations in signal size since the MPV of the charge distribution is uniform along $z$ at $y=\SI{27.5}{\micro\meter}$ as can be seen in \fig{intrapixelChargeMpvGrazing}. Instead, it is probably due to a difference in charge carrier velocity. For particles crossing the pixel close to the implant, the induced signal is mostly due to holes moving away from the pixel implant. Electrons, which have a higher drift velocity, will contribute to the signal when the particle crosses the pixel further away from the implant (towards the $z$-centre), resulting in a faster signal. As the particle crossing point moves even closer to the backside of the sensor, the relative delay increases again because the weighting field is lower near the back electrode (since the pixel does not have an ideal parallel plate geometry), and hence the charge has to drift for some time before significant induction appears as can be seen from the weighting potential shown in \fig{thplWeightingPotential}.
	
\begin{figure}[htbp]
		\centering
		\includegraphics{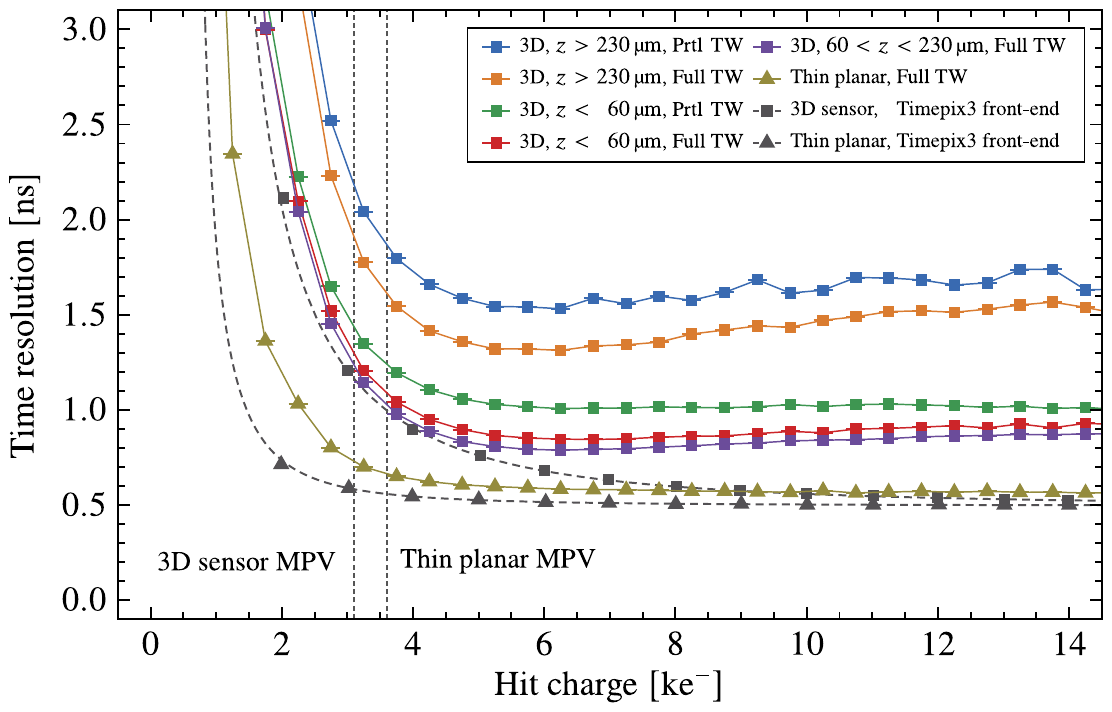}%
		\caption{The time resolution as a function of hit charge for both DUTs. The contribution of the Timepix3 front-end obtained from test pulses is also shown.}
		\label{fig:timeResolutionGrazing}
	\end{figure}

	The hit time resolution is shown in \fig{timeResolutionGrazing} for both DUTs. It can be seen that the 3D detector has the best time resolution in the region $\SI{60}{\micro\metre}<z<\SI{230}{\micro\metre}$. A full timewalk correction improves the time resolution outside of this region because it corrects for the differences in time delay that are not only due to signal size variation. For the region $\SI{60}{\micro\metre}<z<\SI{230}{\micro\metre}$, only the time resolution after a full timewalk correction is shown as the improvement over a partial timewalk correction is not clearly visible in this plot. At the most probable signal charge, the thin planar detector has a better time resolution than the 3D detector because the latter suffers more from jitter in the analog front-end due to a higher pixel capacitance.

	\subsection{Multi-hit time resolution}
\begin{figure}[htbp]
		\centering
		\includegraphics{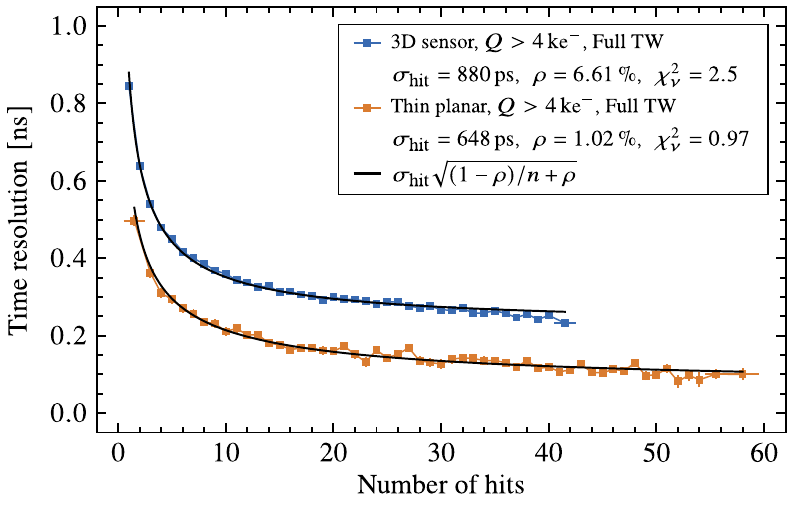}%
		\caption{The cluster time resolution for both DUTs as a function of the number of hits used to calculate the cluster time.}
		\label{fig:multiHitimeResolutionGrazing}
	\end{figure}

	A more precise time measurement might be obtained by combining the single hit time measurements. One way to do this is by calculating a weighted mean of the single hit measurements using the charge dependent time resolution (\fig{timeResolutionGrazing}) in determining the weights as $\sigma_{\textnormal{t}}^{-2}$. The result is shown for both DUTs in \fig{multiHitimeResolutionGrazing}. For the 3D sensor only hits in the region $\SI{60}{\micro\metre}<z<\SI{230}{\micro\metre}$ are used in the weighted mean. The dependence of the cluster time resolution on the number of hits $n$ is modelled as
	\begin{equation}
		\sigma_{\textnormal{cl}}^2\left(n\right) = \left(\frac{1-\rho}{n} + \rho \right) \sigma_{\textnormal{hit}}^2
	\end{equation}
	where $\rho$ can be thought of as the mean correlation between the hits. This expression is exact for performing $n$ measurements of equal time resolution $\sigma_{\textnormal{hit}}$ that are all correlated to each other by the same amount $\rho$. For a weighted mean as performed here, it becomes an approximation.

	The thin planar detector achieves a time resolution of about \SI{100}{\pico\second} for about \num{50} to \num{60} hits, and the 3D detector achieves a resolution of about \SI{250}{\pico\second} for about \num{40} hits. It can be seen that the measurements on both DUTs are not completely uncorrelated since they do not scale as $1/\sqrt{n}$. In principle, correlation between hits is expected since all measurements are performed with a single clock. However, variation in the TDC bin size and time offsets between pixels effectively misalign the TDC bins and act as a type of dither resulting in much less correlation than what would be expected for perfectly aligned TDC bins. Correlations due to clock jitter, on the other hand, are not reduced by this misalignment. The 3D detector suffers more from correlations between the hits than the thin planar detector. This could be because the thin planar detector has more variation in the size of its time bins (\fig{timeBinWidths}).	Detailed investigation of these remaining correlations is outside the scope of this paper, but could be interesting for applications that use ASICs from the Timepix family as a readout for microchannel plates \cite{Tremsin:2020}.

	\section{Conclusion}
	It has been shown that the 3D and thin planar sensors exhibit significant variations in the relative time delay as a function of intrapixel position for particles crossing the detector at both perpendicular and grazing angles of incidence. These variations were corrected by a conventional timewalk correction based on the amount of charge collected in each hit. Using the track information to correct for time variations that are due to spatial dependence of the signal induction offers only about a \SI{1}{\percent} improvement in the overall time resolution in measurements performed at perpendicular incidence. However, the overall time resolution is dominated by the Timepix3 front-end. For the most probable signal charge in a single hit, the time resolution of the 3D detector is dominated by the TDC in the digital front-end. For the thin planar detector, jitter in the analog front-end also has a significant contribution in addition to the TDC. Using the track information to correct for time variations might be useful in a future 4D tracker that uses a faster front-end with a more precise TDC.
	
	At perpendicular incidence the 3D detector suffers least from timewalk because the 3D sensor generates signals that are about a factor six bigger than that of the thin planar sensor due to the difference in thickness. Small signals in the analog front-end of Timepix3 do not only result in an increased time delay, but the timing jitter also increases significantly because the preamplifier output signal crosses the threshold value of the discriminator with a lower slew rate. In spite of this, the thin planar detector has a better time resolution for signals of less than about $\SI{16}{\kilo\electron}$ because the 3D detector has a worse time resolution for particles crossing the electrode regions and it has more analog front-end jitter due to a higher pixel capacitance. Still, at their typical signal size, the 3D detector achieves a better overall time resolution of \SI{567(6)}{\pico\second} compared to \SI{683(8)}{\pico\second} for the thin planar detector.

	Using a grazing angle beam showed that there is a slow region near the backside of the 3D sensor which is probably due to this region not being depleted. The best time resolution for the grazing angle measurements was achieved by the thin planar detector because its lower pixel capacitance gives it an advantage over the 3D detector for small signals. Furthermore, for the thin planar sensor it has been shown that, in terms of relative delay, there is an optimum region in the middle of a pixel cell due to the difference in charge carrier velocity. It would be interesting to compare the thin planar n-in-p sensor that was used in this study to a p-in-n sensor because it can be expected that a p-in-n sensor will have faster signals for charge generated in the region close to the pixel implant since electrons would dominate in the signal induction instead of holes. 

	The possibility of achieving a better time resolution by combining several hits of the same cluster was briefly explored. Although this can indeed give a better time resolution, there is a difference between the DUTs in the amount of correlation in the hit time measurements. To achieve a good time resolution it is vital to have minimum correlations, and a more careful analysis is therefore required to understand the cause of this difference.

	\acknowledgments
	We express our gratitude to the CERN Linear Collider Detector group for providing us with a Timepix3 assembly with an active-edge sensor, and also to Wiktor Byczynski and Raphael Dumps at CERN for their vital support during the test beam period. We also thank our colleagues in the CERN accelerator departments for the excellent performance of the SPS. This research was funded by the Dutch Research Council~(NWO).
	
	\bibliography{bibliography}
	
\end{document}